\documentclass[11pt]{amsart}
\usepackage[russian,english]{babel}
\usepackage{changepage}
\usepackage{cite}
\usepackage[usenames]{color}
\usepackage{comment}
\usepackage[shortlabels]{enumitem}
  \setlist[enumerate,1]{leftmargin=5pt}
  \setlist[itemize,1]{leftmargin=20pt}
  \setlist[description,1]{leftmargin=15pt}
\usepackage{float}\restylefloat{figure}
\usepackage{graphicx}
\usepackage{hyperref}
\usepackage{multicol}
\usepackage{xurl}
\newtheorem{definition}{Definition}

\renewcommand\phi{\varphi}
\newcommand\qef{\hfill$\triangleleft$} %Quod erat faciendum
\newcommand\ru[1]
  {\begin{otherlanguage}{russian}#1\end{otherlanguage}}

\title[Logical foundations]{Logical foundations: \\
\small Personal perspective}
\author{Yuri Gurevich}

\begin{document}
\begin{abstract}
This is an attempt to illustrate the glorious history of logical foundations and to discuss the uncertain future.
\end{abstract}
\maketitle
\thispagestyle{empty}

\subsection*{Apologia\protect\footnote{The article was written for the February 2021 issue of the Logic in Computer Science column in the Bulletin of the European Association for Theoretical Computer Science and is published here as is.}}
I dread this scenario. The deadline is close. A reminder from the Chief Editor arrives.
I forward it to the guest author and find out that they had a difficult real-world problem, and the scheduled article is not ready. I am full of sympathy.
But the result is all the same: the article is not there.

That happened once before in the long history of this column, in 2016.
Then the jubilee of the 1966 Congress of Mathematicians gave me an excuse for a micro-memoir on the subject.

This time around I decided to repurpose a recent talk of mine \cite{BS}. But a talk, especially one with ample time for a subsequent discussion, is much different from a paper. It would normally take me months to turn the talk into a paper. Quick repurposing is necessarily imperfect, to say the least. Hence this apologia.

\section{Prelude}
\label{prelude}

What should logic study?
Of course logic should study deductive reasoning where the conclusions are true whenever the premises are true.
But what other kinds of reasoning and argumentation should logic study? Experts disagree about that. Typically it is required that reasoning be correct in some objective sense, so that, for example, demagoguery is not a subject of logic.

My view of logic is more expansive. Logic should study argumentation of any kind, whether it is directed to a narrow circle of mathematicians, the twelve members of a jury, your family, your government, the voters of your country, the whole humanity. There are many ways to argue. And, yes, demagoguery is one of them, and it should be a subject of logical analysis as well.

\newpage
Look at economics. They used to study almost exclusively the behavior of \emph{homo economicus}, a species of rational agents acting in accordance with their self-interest.
But not any more. Of course economists always knew that people are more complex, and homo economicus is still a useful simplifying assumption. But there are more and more cases where the assumption is not satisfactory any more.

Somewhat similarly, logic studies almost exclusively the reasoning of \emph{homo logicus}, a species of tireless rational agents performing impeccable deduction. Essentially the logic of homo logicus is mathematical logic, and of course mathematical logic is super important.
But people are not homo logicus. Other kinds of argumentation, e.g.\ legal reasoning or political propaganda, are also important and should be studied.
Deductive logic is insufficient even for natural sciences, where inductive reasoning is unavoidable.
In our time, social sciences are coming of age. They study real people, not homo logicus.

Now lets turn attention to logical foundations. I use the term as it is used in ``logical foundations for mathematics'' or ``logical foundations for software,'' and for brevity will often abbreviate the term to just ``foundations.''
Logical foundations for mathematics is a long-established discipline. Logical foundations for software is a younger but recognized discipline. In our opinion, logical foundations of any science should be developed.

Obviously science needs good foundations.
``When the roots are deep,'' says a Chinese proverb, ``there is no reason to fear the wind.''
And sciences should have useful applications, though it is foolish to expect every scientific advance to have immediate applications. Still, as Louis Pasteur said, sciences and the application of science are bound together as a tree and the fruit it bears.
A Russian philosopher-humorist, Michael Zhvanetsky, put it more bluntly: ``It's not enough to know your worth; you still need to be in demand\footnotemark.''
\footnotetext{\ru{Мало знать себе цену; надо ещё пользоваться спросом.}}

But what is the scope of foundations? That aspect is more controversial. As in the case of logic, I take an expansive view in this exposition. If you find this provocative, you are not wrong. It is intended to be so. The goal is to spur a discussion on the role of logical foundations.

While logic and logical foundations are closely related, let's separate concerns. The focus of this paper is logic foundations.
The issue of foundations richly deserves that we
\begin{enumerate}
\item[-]  explore the glorious past,
\item[-]  examine the present, arguably less glorious, and
\item[-]  discuss the uncertain future.
\end{enumerate}
But time is short and my scholarship is limited, so I will only scratch the surface.

\section{Personal experience}

It may be useful to understand where the author is coming from. So let me say a few words about my personal experience with foundations.
The first relevant episode that I remember happened in my middle school, on the outskirts of the industrial Russian city of Chelyabinsk, far from any centers of intellectual life.
(My parents were peasants \citen{WikA}
who moved during the collectivization \citen{WikB}
to a city and to the proletariat.)

The episode occurred during a lesson on trigonometry.
The teacher was proving the congruence of two triangles. Not many students followed or even listened.
``Let's take a special third triangle,'' she said. I raised my hand: ``Where from?''
For a moment, the teacher looked confused, and now everybody looked at her.
``Shut up!'' she replied\footnote{\ru{Заткнись!}}.
But my question was sincere. I did not intend to give her trouble. We were accustomed to shortages. Why should there be abundance of triangles?
Apparently, basic mathematical intuition is not prior knowledge as some philosophers surmised; it needs to be learned.

We didn't study mathematical analysis in high school but I participated in a mathematical olympiad and received a bunch of mathematical books, including Khinchin's ``Short course in mathematical analysis'' \cite{Khinchin}.
The book was well-written and answered some questions that bothered me. It also contributed to something else.
In 1957, I entered the local Polytechnic where they did teach us calculus. Sometimes the alleged proofs of my professor were not real proofs. Whenever I attempted to point this out, he made everybody laugh at me.

In the middle of the 1958-59 academic year, I succeeded to transfer to the Ural State University in Sverdlovsk (now Yekaterinburg) where my professors knew their stuff.
In my 1962 diploma thesis, I solved an open problem in abstract group theory \cite{G001}; an ad hoc construction produced the desired counterexample but let me somewhat dissatisfied.
The issue seemed divorced from the beautiful classical mathematics that we studied.
At that moment of my little crisis, I got a birthday present that changed my career:  Kleene's ``Introduction to metamathematics,''
a powerful exposition of the foundations of mathematics,
skillfully translated and amply commented by Alexander  Esenin-Volpin  \cite{Kleene}.

I read the book as if it was an engaging detective story, and now I wanted to be a logician. But there were no logicians in the Urals. Formal logic was, for all practical purposes, forbidden in Stalin's time, and it was recovering slowly. To put my foot in the door, I worked on the stuff of interest to the Novosibirsk ``Algebra and Logic'' seminar headed by Academician Anatoly I. Maltsev who started as a logician and returned to logic when it was safe to do so.

In 1973 my family was allowed to leave the ``land of victorious socialism" for Israel.
I intended to transit to computer science, but the Jerusalem logic seminars proved to be too attractive. I couldn't miss the opportunity to study modern logic and collaborate with Saharon Shelah, the world's best model theorist. Eventually the project that Shelah and I were working upon grew close to completion, and I sought ways to transit to computer science.
That brought me to the University of Michigan where I taught 1982--98 and where I zeroed in on the foundational problem what an algorithm is.
I came up with abstract state machines (ASMs) and a thesis that, semantically, algorithms are ASMs. In 1998, Microsoft Research Redmond invited me to start a group on Foundations of Software Engineering and build an ASM-based tool for software specification, verification, testing.

I worked at Microsoft for 20 years, 1998--2018. Professionally, these were the most satisfying years of my life though there were many challenges.
To begin with, it was hard for this mathematician, who programmed only one of the first Soviet computers, to hire right people. But it worked. We built an ASM-based tool, Spec Explorer \cite{WikC},
which was adopted by the Windows team. In 2003, I asked my first hire, Wolfram Schulte, to assume the management of the group, and I was free to move on research-wise.

I had an opportunity to work with Microsoft groups on various issues including access control, cybersecurity, privacy, and quantum computing.
Foundational problems arose in almost all cases. Here is but one example. When we worked on the ``Inverse privacy'' \cite{G222}, it became clear that the logic foundations of privacy are all but nonexistent. Experts disagree on what privacy is. In \cite[\S2]{G222}, we started a foundational  development but carried it only to the small extent sufficient for our needs in that article.

\section{The glorious past}

I give just a few examples of past foundational advances. The examples are listed in chronological order; otherwise there is no pretense to be systematic.

One can speak about logical foundations of mathematics and logical foundations of natural sciences, but we start with logical foundations of human civilization itself.

\bigskip\noindent
\textbf{\large Invention of alphabet.}\mbox{}

People make a continuum of sounds, and the range of sounds varies from one language to another, from one dialect to another, and even from one person to another.
Recall the biblical ``shibboleth'' story in this connection.
About 1800 BC, the first alphabet was invented.
The continuum of sounds was reduced to just a few, twenty two to be exact.
It seems miraculous that the alphabetic principle was conceived, implemented, and accepted.

We speak here about consonant sounds only because vowels play an auxiliary role in Semitic languages and the inventors spoke a Semitic language. Later the Greeks modified the Semitic alphabet into one that includes consonants and vowels.

\begin{center}
\includegraphics[scale = .4]{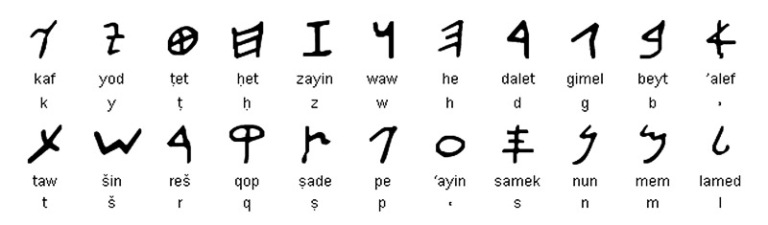}\\
\textbf{Canaanite Alphabet
\footnote{Picture from https://sinangharaibeh.wordpress.com/2012/04/04/414/}}
\end{center}

Did the alphabetic principle somehow evolve from earlier writing systems or did the idea occur to a single individual?
According to the most popular theory, the first alphabet was not entirely invented but partially discovered.  Egyptian hieroglyphs combined logographic syllabic elements and elements representing consonants, and so ``Egyptian’s logo-consonantal system set the stage for purely consonantal systems'' \cite[p.~22]{Hoffman}.

Notice a logical angle. A written word usually does not reflect its meaning, the thing that it denotes. Instead, it represents the sounds in the name of the thing. In that sense, alphabetic writing is a triumph of syntax over semantics.

Can a civilization get by without an alphabet. Yes, think Chinese civilization. But there is a price. For example, printing was known in China centuries before Gutenberg, but in the absence of an alphabet, printing wasn't that useful. \qef

\noindent
\textbf{\large What is knowledge?}\mbox{}

\smallskip
In September 1995 I gave a few lectures at the University of Amsterdam. After the first lecture, a young Russian woman (Natasha Alechina, as I learned later) asked me whether I fashion my BEATCS dialogs after Plato.
Actually, I had never read Plato. At the second lecture, she gave me a copy of Plato's Theaetetus. To say that I was impressed by this dialog would be an understatement.

\begin{multicols}{2}
In the dialog, written about 369 BC, Socrates asks a young mathematician Theaetetus what knowledge is.  Theaetetus has no idea.  Socrates prods him. Theaetetus comes up with a definition that Socrates dismisses out of hand. Theaetetus thinks hard and comes up with
\begin{definition}
Knowledge is perception.
\end{definition}
\noindent
which Socrates finds worthy of consideration.

\begin{center}
\includegraphics[scale = .3]{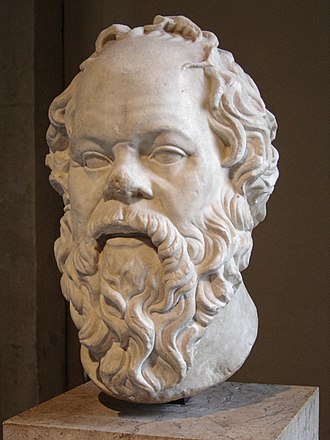}\\
\small A marble head of Socrates\\ in the Louvre
(from Wikipedia)
\end{center}
\end{multicols}

A scholarly discussion ensues.  Socrates analyzes the definition, quotes other philosophers and criticizes them. In the end, Definition~1 is rejected. Socrates prods Theaetetus to think even harder, and eventually Theaetetus improves upon his definition.
\begin{definition}
Knowledge is true belief.
\end{definition}
Socrates likes the new definition better but again, after a profound discussion, he rejects it. Like a midwife, Socrates tries to help Theaetetus to birth the right definition. Theaetetus comes up with
\begin{definition}
Knowledge is true belief with an account.
\end{definition}
Socrates likes the new definition better yet, but again, after another profound discussion, he rejects it. ``Let's meet in the morning and continue,'' he says.

And that is where we are today, more or less, except that, instead of ``true belief with an account,'' we say ``justified true belief.'' \qef

\noindent
\textbf{\large Axiomatic method}\mbox{}

The axiomatic method appeared first in Euclid's geometry as far as we know. Euclid lived in Alexandria from mid-fourth century to mid-third century BC.
His geometry, taught from his days to ours, is one of the top achievements of Hellenistic mathematics. According to Lucio Russo \cite{Russo}, science did not start in the Renaissance, but flourished already in the Hellenistic period, whose golden age was
from the late fourth to the late second century BC. The Hellenistic period ended with the annexation of Egypt by Rome in 30 BC, and science did not survive for long after that. \qef

\bigskip\noindent
\textbf{\large Symbolic notation}\mbox{}

A systematic use of symbolic notation, like letters for constants and unknowns, is most useful and seems very natural, but the history of symbolic notation is long and involved \cite{Katz}. Here we'll say only a few words.

\hspace{-20pt}
\begin{minipage}{0.75\textwidth}
Fran\c cois Vi\`ete (1540--1603) was French and published under the Latin version of his name, Franciscus Vieta.
In the 1591 ``Introduction to the analytic arts'' he used capital vowels for unknowns and capital consonants for constants.
``If this seems reminiscent in principle of our modern notation of $x, y$, and $z$ for unknowns and $a, b, c$, etc. for indeterminate magnitudes, a convention which we owe to René Descartes in the seventeenth century \dots, it is important to recognize that Viète’s symbols or `species,' unlike ours, carried explicit geometrical meaning. They had dimension, and only expressions of the same dimension were commensurate'' \cite[p.~316]{Katz}. \qef
\end{minipage}\hfill
\begin{minipage}{0.2\textwidth}
\includegraphics[scale = .3]{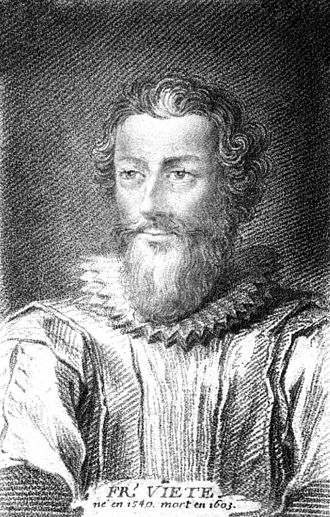}\\ \small
Fran\c cois Vi\`ete\\
(from Wikipedia)
\end{minipage}

\bigskip\noindent
\textbf{\large Infinitesimal calculus}\mbox{}

The history and prehistory of the infinitesimal calculus, from antiquity (yes, from antiquity \cite{Bell}) to our days is rich and involves many important foundational ideas.

Gottfried Wilhelm Leibniz (1646--1716), one of the two discoverers of the calculus --- the other being Isaac Newton (1643--1727) of course  --- was a philosopher and logician.
``Leibniz himself said, without much exaggeration, that all his mathematical discoveries arose merely from the fact that he succeeded in finding symbols which appropriately expressed quantities and their relations'' \cite[p.~516]{Jourdain}.

But, in addition to these two 17th century heroes of the calculus, there are important 19th century heroes, including Bernard Bolzano (1781-1848), Augustin-Louis Cauchy (1789--1857), %Peter Gustav Lejeune Dirichlet (1805-1859),
Karl Weierstrass (1815-1897), Richard Dedekind (1831-1916), and Georg Cantor (1845 -- 1918). In this connection, let us mention a story, originated in the school of Pythagoras of Samos (c. 570 -- 495 BC) and finally resolved only in 20th century.

The story\footnote{inspired by article \cite{Alexander} where the story is told dramatically but incompletely. (Allegedly that article was published first in \emph{Scientific American} 310:4 April 2014, but the article there has a different content.)}
starts with the discovery that the hypotenuse of an isosceles right triangle is incommensurable with its legs. In modern terms, this means essentially that $\sqrt2$ is irrational.
It is hard to know to what extent the story is factual, but apparently the discovery was made by Hippasus of Metapontum,
who might have been killed by Pythagorean zealots because the discovery contradicted a Pythagorean dictum according to which a small indivisible unit should fit evenly into all three sides of the triangle.

Arguably the problem was solved by Eudoxus's theory of proportions which is presented in Book V of Euclid's ``The Elements'' and which is a precursor of the theory of Dedekind cuts.
%That solution has been somewhat problematic \cite{Ofman}.
The theory of infinitesimals gave a different view on the problem.
An infinitesimal unit seems to fit evenly into every side of any triangle.
But of course infinitesimals themselves are problematic;
that problem was solved in the 19th century with the development of the theory of calculus that did not use infinitesimals.
And the nature of infinitesimals was clarified in the 20th century by nonstandard analysis \cite{Robinson}. No infinitesimal $x$ fits evenly into 1 and $\sqrt2$, but there are nonstandard integers $I$ and $J$ such that the standard parts of products $Ix$ and $Jx$ are 1 and $\sqrt2$ respectively. \qef

\bigskip\noindent
\textbf{\large People}\mbox{}

On different occasions, I had spoken about relatively recent heroes of logical foundations including Frege, Russell, Hilbert and his students, Church, Tarski, and especially G\"odel, Turing and Kolmogorov.

\begin{multicols}{3}
%\noindent
\includegraphics[scale = .27]{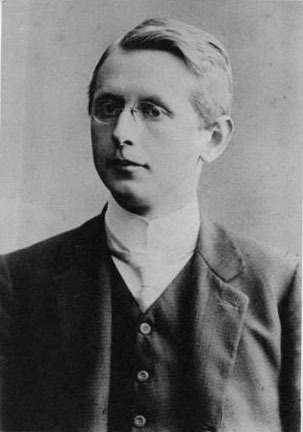}\\
\mbox{}\quad Hermann Weyl

\noindent
In the talk underlying this paper I spoke about Hermann Weyl and John von Neumann, mostly because I encountered them twice,
first in foundations of mathematics and computing, and then, many years later, in foundations of quantum theory.

\hfill
\includegraphics[scale = .25, trim = 0 80 0 0, clip]
{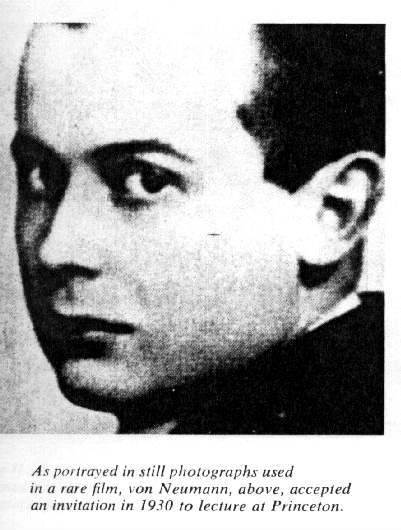}\\
\mbox{}\hfill John von Neumann
\end{multicols}

Young Hermann Weyl was a passionate constructivist, attracted to idealist philosophy, but also a pragmatic and prolific scientist with a taste for foundations.
His view changed over time. ``In his youth he inclined towards the idealism of Kant and Fichte, and later came to be influenced by Husserl’s phenomenology'' \cite[\S5.7]{Bell}.
The choice of epigraph for Weyl's book ``Philosophy of mathematics and natural sciences'' \cite{Weyl} is telling:
\begin{quote}
Home is where one starts from. As we grow older\\
The world becomes stranger, the pattern more complicated\\
Of dead and living.\\[5pt]
\mbox{}\qquad T.S. Eliot, Four Quartets, East Coker, V.
\end{quote}

John von Neumann was a true polymath. Here are just the titles of his six-volume collected works (Pergamon Press, 1961-1963):
\begin{itemize}
\item[1] Logic, theory of sets and quantum mechanics
\item[2] Operators, ergodic theory and almost periodic functions on a group
\item[3] Ring of operators
\item[4] Continuous geometry and other topics
\item[5] Design of computers, theory of automata and numerical analysis
\item[6] Theory of games, astrophysics, hydrodynamics and meteorology
\end{itemize}
Somebody said that, in the day that von Neumann died, twenty people became the top researchers in their fields.

And both, Hermann Weyl and John von Neumann, were human enough to err. Weyl lost a bet about the future of constructivism \cite{G123}. As far as an error in von Neumann's book \cite{JvN},  David Mermin said it well:

\begin{quote}
Many generations of graduate students who might have been tempted to try to construct hidden-variables theories were beaten into submission by the claim that von Neumann \dots\ had proved that it could not be done. A few years later \dots\ Grete Hermann \dots\ pointed out a glaring deficiency in the argument, but she seems to have been entirely ignored. Everybody continued to cite the von Neumann proof. A third of a century passed before John Bell \dots\ rediscovered the fact that von Neumann's no-hidden-variables proof was based on an assumption that can only be described as silly \cite{Mermin}.
\end{quote}

\section{Present}

For a while, after the second world war, logic was popular and foundational. After all, electronic computers were designed on a logic foundation. Here are a few other examples.
\begin{itemize}
\item In emerging artificial intelligence, symbolic approach dominated.
\item Logic programming was quite a fad, partially because of the Japanese national Fifth Generation project 1982-1992.
\item Non-standard analysis arose on logic foundations.
\item Forcing revolutionized set theory.
\end{itemize}

But in the 1980s things started to change.
While the need for logic foundation research never was greater, less attention was given to foundational issues, even in areas like set theory.

By and large, logic groups are slowly fizzling out at top mathematics and philosophy departments.
For example, the Mathematical Department of ETH Zurich that used to have Zermelo, Bernays, and Specker, let its last three logicians retire without hiring a single new logician.

There are many logicians in computer science but not much foundational work is done there either. In particular, the golden age of logic in artificial intelligence is behind us.

\section{Future}

Scientific progress will continue. Foundational problems will inevitably arise and will be addressed. But what role will logicians play?

The future may be bleak for logicians.
Foundational logic research fades away.  Significant logic areas become parts of mathematics or computer science.

\begin{center}
\begin{minipage}{.15\textwidth}
\phantom{XXXXXXXXXXXX}
\end{minipage}
\begin{minipage}{.6\textwidth}
Between the potency\\
And the existence\\
Between the essence\\
And the descent\\
Falls the Shadow\\
$\vdots$ \\
This is the way the world ends\\
Not with a bang but a whimper.\\[5pt]
\mbox{}\qquad T.S. Eliot, The Hollow Men
\end{minipage}
\end{center}

\bigskip
But this does not have to be so.
The future could be bright, albeit challenging.
First of all, I think, we need to discuss the issue explicitly. This paper is an attempt to spur such a discussion.
One way or another, logic research should recover its foundational spirit.

\medskip\noindent
\texttt{Make logic research more comprehensive.}\\[3pt]
You may buy into this idea, even if my view of logic seems too expansive to you. To recall, my view is that logic is the science of reasoning and argument, which does not reduce to the logic of mathematics or even the logic of science.

\medskip\noindent
\texttt{Make foundational research more comprehensive.}\\[3pt]
Throughout ages, the logic way of thinking made great foundational contributions to science and civilization itself. We should revive that tradition.

\smallskip
There are many foundational problems to explore.
In the rest of this section, I give a few rather general examples.
The list is not the result of a careful investigation.
These are the examples that occurred to me as I was preparing the talk underlying this paper. There was a good hour-long discussion after the talk, but feedback on the examples was limited.

\bigskip\noindent
\textbf{\large Inductive inference.}

Already Aristotle mentioned inductive inference in addition to deductive. Yet, inductive inference remains a challenge.

The problem has been addressed by philosophers, notably David Hume (1711--1776) and Karl Popper (1902--1994). Popper's falsifiability principle has been influential with natural scientists. Critics of Popper point out that his principle is simplistic for natural sciences.
Pragmatic natural scientists use Popper's principle as a general guide, not literally.

\texttt{An aside.}
If a scientific theory $T$ has the form $\forall x \phi(x)$, where the range of $x$ is infinite and $\phi$ is experimentally decidable, then indeed $T$ is experimentally unverifiable, but a single counterexample falsifies $T$.
If $T$ has the form $\exists x \phi(x)$, then $T$ is experimentally unfalsifiable but a single example of $\phi$ verifies $T$. If the form of $T$ is more complicated, say $\forall x \exists y \psi(x,y)$, then $T$ is neither falsifiable nor verifiable experimentally. (This is not my observation, but I do not have a reference.)

Bayesian inference is explicitly pragmatic. You don't aim to decide whether the hypothesis $H$ in question is true or false. Instead you aim just to confirm or disconfirm $H$ to the extent of the available evidence. Bayesian inference is a useful and widely used tool. \qef

\bigskip\noindent
\textbf{\large Information, knowledge, privacy}

What is information? What is algebra of information? What is logic of information? My coauthors and I attempted to address the issue in article \cite{G198} and in subsequent articles on infon logic.

The same questions arise for knowledge and for privacy instead of information. As far as knowledge is concerned, we have more data than Plato did but only limited progress so far. \qef

\bigskip\noindent
\textbf{\large Life sciences}

Neuroscience is one of the most fascinating and fast developing sciences. Can foundational logic research be useful in neuroscience and life sciences in general? This is hard to tell.
``If people do not believe that mathematics is simple,'' wrote John von Neumann, ``it is only because they do not realize how complicated life is."
But we will not know the answer to the question above if we don't try. One more specific question is related to Daniel Kahneman's ``Thinking fast and slow'' \cite{Kahneman}.
The slow thinking is the conscious thinking of homo sapiens. The fast thinking is the unconscious thinking that homo sapiens inherited from preceding species. What is the logic of fast thinking?
It is much different from Bayesian inference as Kahneman convincingly argues. \qef

\bigskip\noindent
\textbf{\large Social sciences}

Social sciences are becoming, and in some cases have become real sciences \cite{Watts}. They may be even more involved than life sciences. Can foundational logic research be useful there? I don't see why not. We certainly should try.
For example, can one objectively define the degrees of spin and the fakeness of news?

Compared to legal reasoning, mathematical reasoning is easy. Outside mathematics, human pronouncements may not be --- and usually are not --- either true or false.
A sentence, uttered by a person $A$ to people $B$ and $C$ may mean different things to $B$, to $C$, and to $A$.

\begin{quote}
What is needed in law, if law is to become more scientific in the future than it has been in the past, is a body of learning from which we can predict that what looks like a straight story or a straight sale from one standpoint will look like a crooked story or a crooked sale from another, and from which we can predict the successive ``distortions" that any observed social fact will undergo as it passes through different value-charged fields in the ``world-line" of its history \cite[p.~243]{Cohen}.
\end{quote}

Of course such predictions would be of various degrees of confidence.
It would be more useful to reason with predictions. \qef

\begin{comment}
\bigskip
Back to philosophy?

To an extent yes, but:
\begin{itemize}
\item Enable the virtuous circle of incremental-improvement.\\
{\small\quad Recall Thomas Kuhn's story on electricity.}
\item Have applications in mind.\\
{\small\quad Recall how the concept of set became prevalent in mathematics, how Kolmogorov's axiomatization of probabilities succeeded.}
\item Prove theorems.
\end{itemize}
\end{comment}

\section*{Summary}

I attempted to contrast the glorious past of logic foundations with less glorious present and very uncertain future. I treated foundations rather generally. But the picture is similar if one restricts attention to foundations of mathematics. In the first part of the 20th century mathematical logic was prestigious and attracted attention of giants like David Hilbert, John von Neumann, Andrey Kolmogorov, and Alan Turing.
Today, mathematical logic is a respectable discipline but, in many departments, logicians need to justify their field. Are we observing a natural aging of the science of logic?
Or can we rekindle the foundational spirit of logic? Does logic have the potential to regain its prestige, at least partially? These questions call for self-examination and discussion.

\subsection*{Acknowledgment}
Many thanks to Andreas Blass for useful discussions and sanity check.

\end{document}